# *ab*-plane tilt angles in REBCO conductors

Jun Lu, Yan Xin, Yifei Zhang, and Hongyu Bai

***Abstract*—Critical current ($I_c$) of REBCO tapes is strongly anisotropic with respect to the orientation of the magnetic field. Usually, $I_c$ is at maximum when the *ab*-plane of the REBCO crystal is parallel to the magnetic field. In commercial REBCO tapes, it is commonly assumed that the *ab*-plane is coincide with the tape plane. While in fact, the *ab*-plane is near but slightly tilted from the tape plane in the transverse direction. To accurately measure $I_c$ as a function of the field angle $\theta$, which is defined as the angle between ab-plane and the magnetic field direction, and to design and fabricate REBCO magnet coils based on the measured $I_c(\theta)$, it is important to measure the tilt angle. In this work, we used x-ray diffraction (XRD) to measure the tilt angles at room temperature for a large number of REBCO conductors made by SuperPower Inc. Transmission electron microscopy (TEM) was also used to investigate the origin of this tilt. The measured data are presented, and the measurement uncertainty is discussed.**

## I. Introduction

AFTER successful development of the 32 T all-superconducting magnet using REBCO double-pancake technology at the National High Magnetic Field Laboratory (NHMFL) [1], the development of a 40 T all-superconducting magnet is under way [2]. As a part of comprehensive REBCO conductor characterization program for the 40 T project, critical current ($I_c$) of every incoming REBCO tape is measured at 4.2 K in magnetic field between 7 and 15 T, and at 18° field angle which is defined as the angle between magnetic field direction and the REBCO crystal *ab*-plane (18° is a reference angle which is the maximum field angle of the 32 T magnet) [3]. It was found that a 180° in-plane rotation of a sample resulted in a significantly different $I_c$ value. The difference was as much as 20% of the $I_c$. This can be explained as the follows.

The $I_c$ of a REBCO tape is strongly dependent on the field angle $\theta$. $I_c(\theta)$ is usually at maximum when $\theta = 0$, i.e., magnetic field is parallel to REBCO *ab*-plane, due to the intrinsic pinning effect and electron mass anisotropy. In $I_c(\theta)$ measurement and REBCO magnet design, it is commonly assumed that the *ab*-plane coincide with the geometric plane of the REBCO tape. In most commercial REBCO tapes, however, the REBCO *ab*-plane is slightly tilted with respect to the tape geometric plane [4]-[9]. For a tilt angle $\beta$, the actual field angle $\theta = \theta_0 \pm \beta$, where $\theta_0$ is nominal field angle which is the angle between magnetic field and the tape plane. Evidently the above mentioned two different $I_c$ values measured at nominal 18° correspond to $I_c$ (18° + $\beta$) and $I_c$ (18° - $\beta$) as depicted in Fig. 1. For each $I_c(\theta)$ measurement, it is important to know the direction and magnitude of the tilt. So the true $\theta$ values are documented in $I_c(\theta)$ database which will be the base for magnet design calculations. Moreover, in the process of fabricating REBCO magnet coils, it is possible to lose the track of the tilt direction. Therefore it is

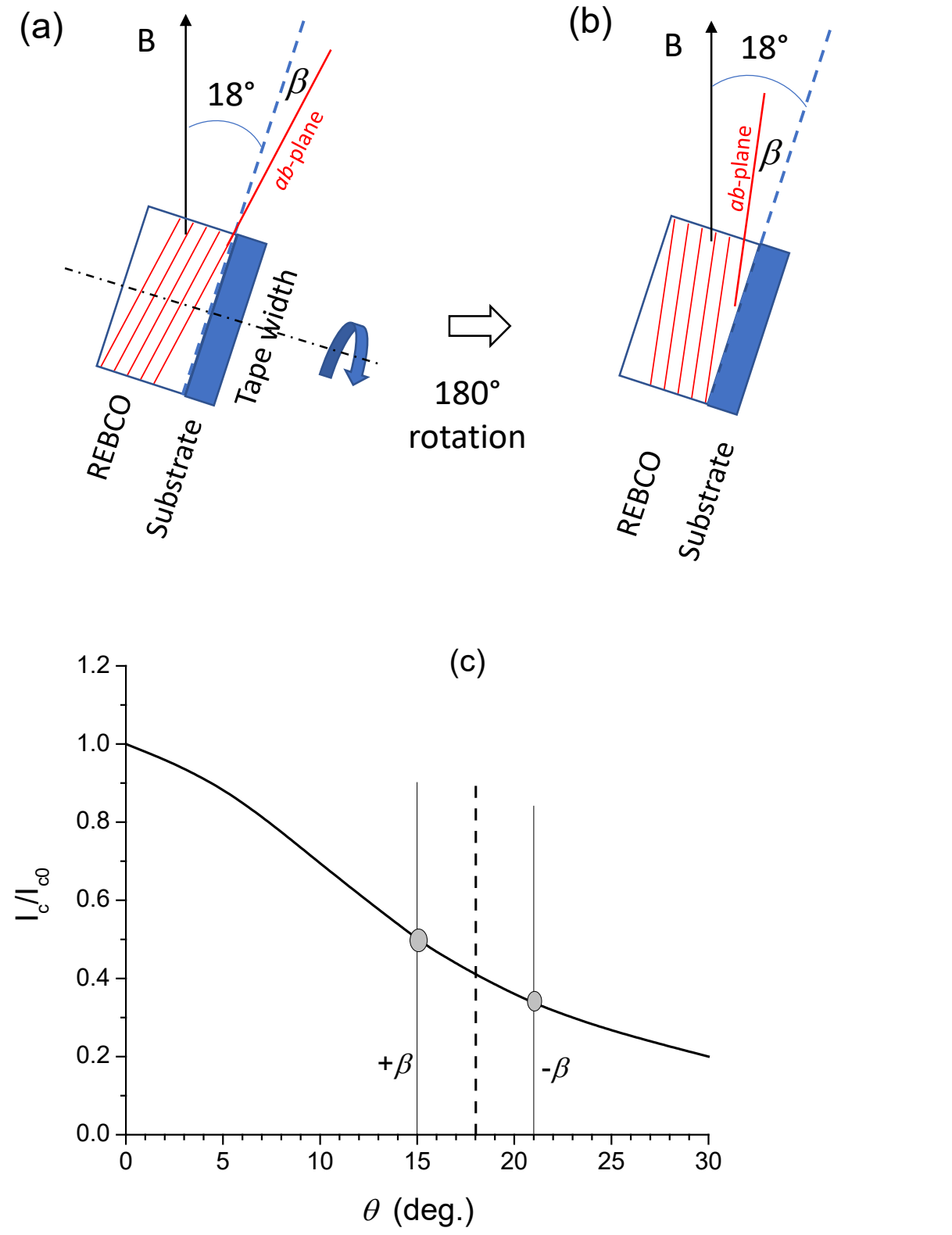

Fig. 1. Schematic showing the effect of *ab*-plane tilt angle $\beta$. (a) $\theta = 18 + \beta$ due to *ab*-plane tilt. (b) after 180° in-plane rotation, $\theta = 18 - \beta$. (c) the significant impact of the tilt polarity on $I_c$, where $\theta$ is the angle between the field and the *ab*-plane.

Manuscript receipt and acceptance dates will be inserted here. This work was performed at the National High Magnetic Field Laboratory, which is supported by National Science Foundation Cooperative Agreement No. DMR-1644779, DMR-1839796, DMR- 2131790, and the State of Florida. *(Corresponding author: Jun Lu, junlu@magnet.fsu.edu.)*

Jun Lu, Yan Xin and Hongyu Bai are with the Magnet Science and Technology division, National High Magnetic Laboratory, Tallahassee, FL 32310, USA (e-mail: junlu@magnet.fsu.edu).

Yifei Zhang is with SuperPower Inc., Glenville, NY 12302 USA.

important to experimentally verify that the direction of the tilt $\beta$ for each coil module is consistent with the design that is calculated based on the $I_c(\theta)$ data.

In this work, we developed a method to measure the direction and magnitude of the *ab*-plane tilt angle by using x-ray diffraction (XRD). The measured tilt angles for many SuperPower conductors are presented; and the origin of the tilt angle is discussed.

## II. Experiment

### A. XRD sample preparation

REBCO tapes measured in this work were SCS4050-AP with 20 µm copper stabilizers made by SuperPower Inc. The REBCO layer was grown by MOCVD, on an IBAD buffer layer which is grown on a 50 µm Hastelloy substrate. The buffer layer consists of $LaMnO_3$(LMO)/MgO/$Y_2O_3$/$Al_2O_3$/Hastelloy. The MOCVD REBCO layer was doped with 7.5 at% of Zr to form $BaZrO_3$ (BZO) nanorods as artificial pinning centers. The samples for XRD measurement were 15 mm long with one corner cut to indicate the longitudinal direction of REBCO growth. The copper stabilizer layer was chemically removed by APS-100 copper etchant, leaving only silver protective layer on REBCO. The sample was then glued to a 25 x 75 x 1 $mm^3$ glass slide by Super Glue. Pressure was applied during the glue curing process to ensure a thin, and uniform glue layer. To maximize the XRD signal, some samples had their silver layer also removed with $H_2O_2$ (30 wt%) : $HN_4OH$ (30 wt%) = 2:5 in volume.

### B. XRD and transmission electron microscopy (TEM) measurements

The XRD was performed with Cu Kα radiation in a Rigaku SmartLab diffractometer which was equipped with a four-circle goniometer (the cradle). Various XRD techniques were used to characterize the REBCO tapes. The REBCO (005) rocking curve scan with 2θ = 38.4° was primarily used for accurate *ab*-plane angle measurement. In the rocking curve measurement, REBCO tape longitudinal direction was perpendicular to the diffraction plane, which is the plane of both incident and diffracted x-ray beams; and the REBCO tape plane is coincide with the specimen plane.

The transmission electron microscopy (TEM) was performed to view the transverse cross-section of REBCO tape. A TEM was prepared using focused ion-beam in the Thermal Fisher Scientific Helios G4 scanning electron microscope/focused ion-beam system. The TEM used was the probe aberration-corrected JEOL JEM-ARM200cF operated at 200 kV.

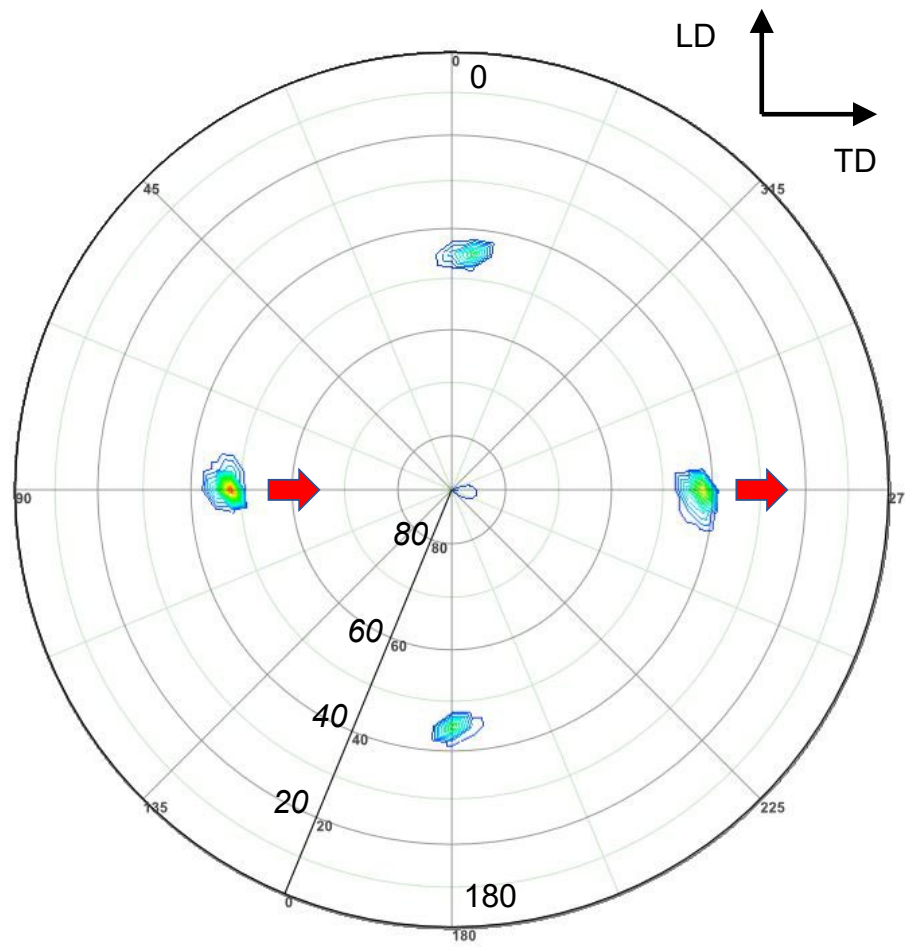

Fig. 2. A XRD χ−ϕ pole figure of REBCO (103) diffractions at $2\theta$= 32.42°. REBCO longitudinal direction (LD) and transverse direction (TD) are indicated. The arrows indicate the tilt in REBCO tape transverse direction by about 3°. There is no tilt in the orthogonal (longitudinal) direction.

## III. Experimental Results

### A. XRD characterization

To identify the direction of the tilt, a XRD $\chi-\phi$ pole figure of REBCO {103} diffractions was obtained, as shown in Fig. 2. All four {103} diffractions are shown clearly. In the tape longitudinal direction (LD), the χ angle of two REBCO {103} peaks are consistent with the calculated value of 46.1°. Whereas the two diffractions in the transverse direction (TD) are shifted from χ = 46.1° by about 3° to lower and higher χ angles respectively as indicated by the arrows. Hence the *ab*-plane tilt was only observed in the transverse direction. It is noticed that the diffraction at $\phi$ = 0° has an unignorable $\phi$ angle offset. The reason for this is unclear.

Before we delve into the detailed analysis of the tilt angle, it is helpful to do a conventional thin film θ-2θ scan to characterize the layered structure in the out-of-plane (*c* axis) direction. For this measurement, the silver layer was chemically removed. After the scan of the sample which consists of both REBCO and buffer layers, the REBCO layer was removed by etching using a diluted $HNO_3$ solution. This way, the diffraction from the buffer layer alone can be shown clearly. Fig. 3 shows the θ-2θ patterns (corrected for the tilt) for (a) the REBCO/buffer layer and (b) the buffer layer only. Diffractions from REBCO (00*l*), as well as MgO (002)/LMO (002) and Hastelloy (220) are clearly observed. A small number of copper oxide and a-axis grain inclusions were observed on REBCO surface by SEM for a number of samples [10], but XRD did not show significant diffractions from them.

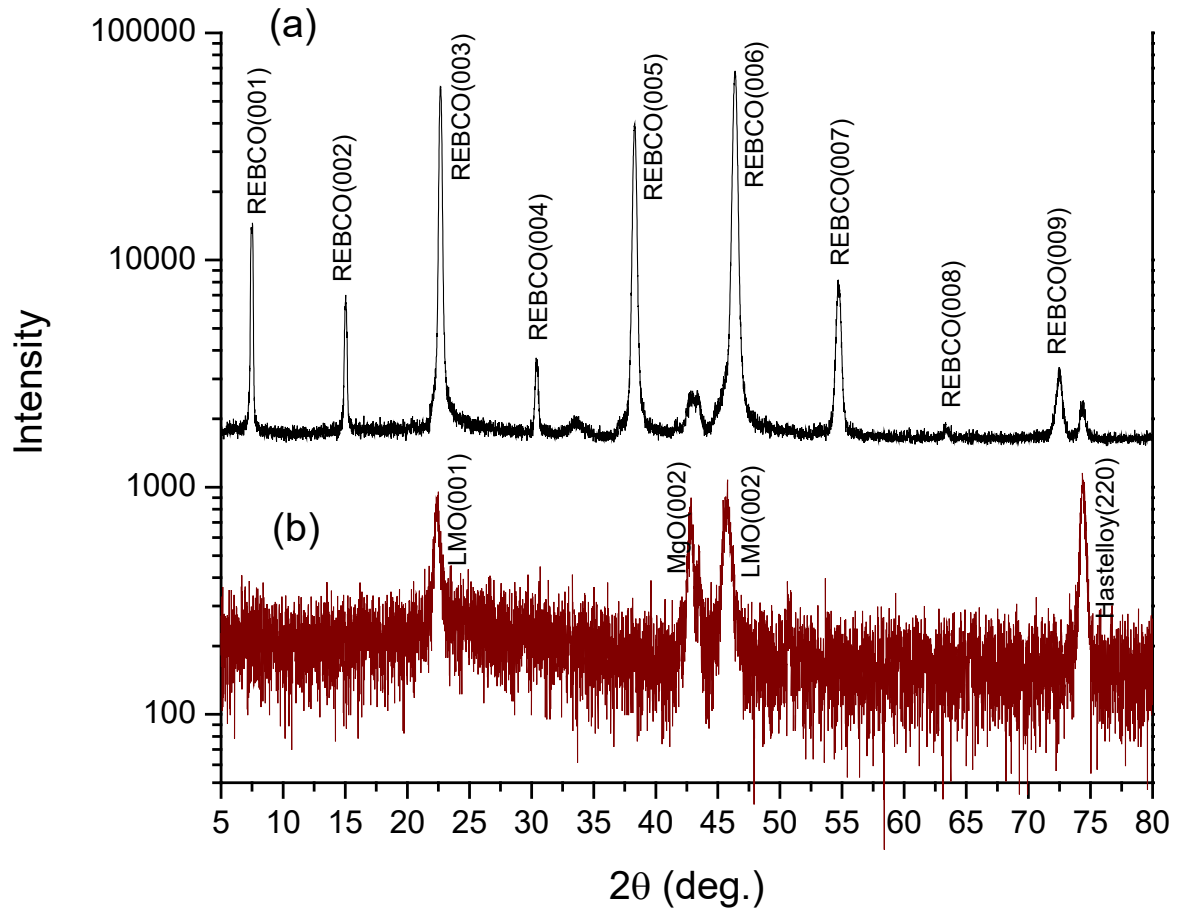

Fig. 3. θ-2θ scan of (a) REBCO film and (b) buffer layer + Hastelloy substrate.

The most efficient method to accurately measure *ab*-plane tilt angle is to perform rocking curve scan (ω scan) of REBCO (005) diffraction. In this measurement, ω axis was scanned while 2θ was set at 38.4°. When *ab*-plane coincides with the tape plane, the nominal peak position is ω = 19.2°. The deviation of the actual peak position from 19.2° is the tilt angle. Fig. 4(a) shows the rocking curves of two samples which tilted about 2° from the nominal position (tape plane) to opposite directions. When the REBCO layer of sample 2 was removed, it was discovered that MgO (002) rocking curve in the buffer layer is also tilted by the same amount from the nominal position of ω = 21.5° (Fig. 4(b)). This strongly suggests that the *ab*-plane tilt is originated from the IBAD-MgO layer, given the fact that the buffer layers before the IBAD-MgO layer are not textured.

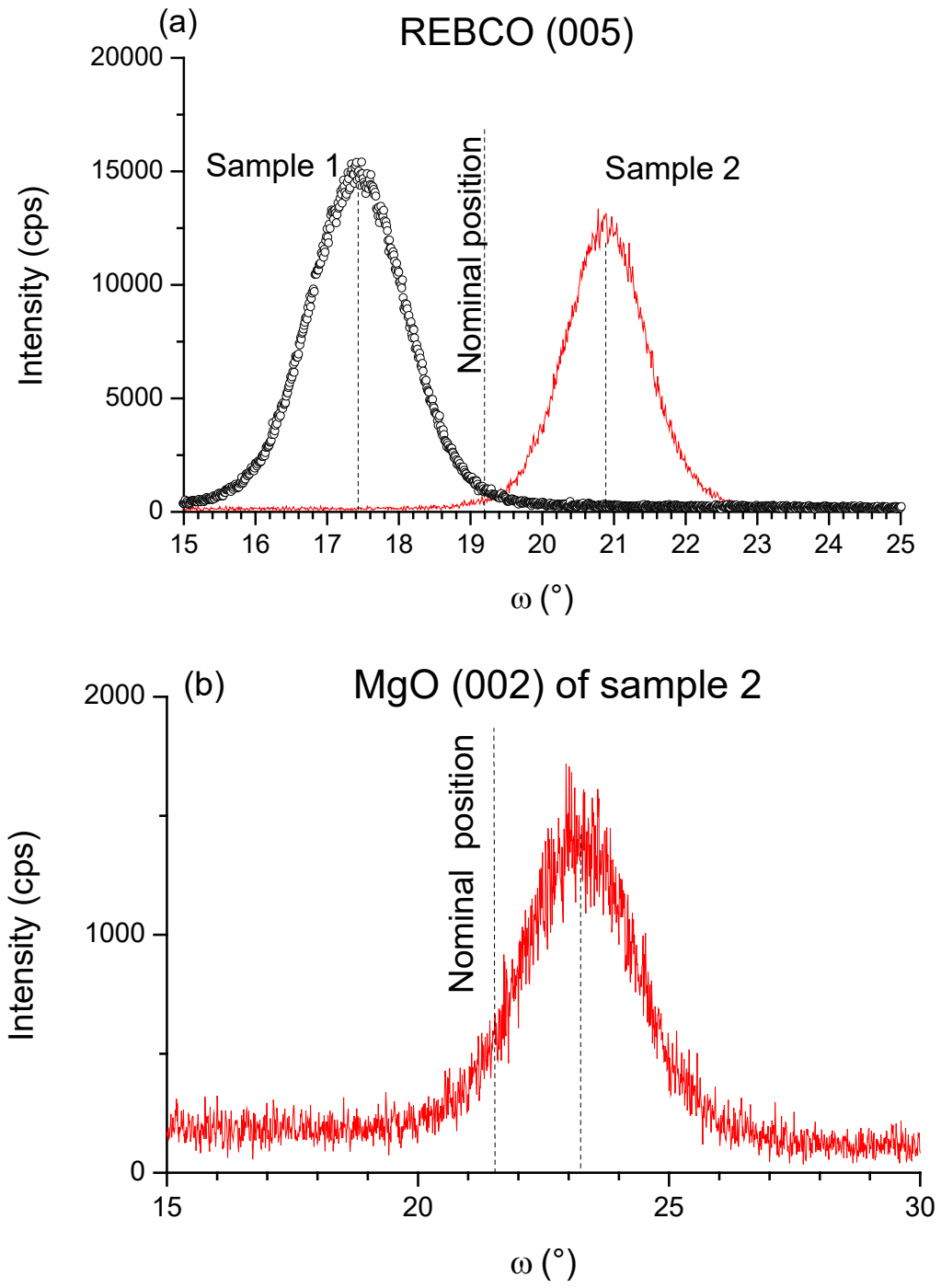

Fig. 4. (a) REBCO (005) rocking curve peaks of two samples tilted about 2° from the nominal position of ω = 19.2° in opposite directions. (b) MgO (002) rocking curve peak of sample 2 is also tilted about 2° in the same direction as its REBCO (005) counterpart.

### B. *Origin of the tilt by high-resolution TEM*

It is imperative to directly image the buffer layer of LMO/MgO/$Y_2O_3$/$Al_2O_3$/Hasterlloy to verify the origin of the tilt as suggested by XRD. Considering the interfaces in the buffer layers are all parallel to the tape face, high-resolution TEM (HRTEM) was performed to image the atomic planes in the buffer layers and measure the tilt angle with respect of the interfaces. HRTEM showed that $Al_2O_3$ and $Y_2O_3$ layers deposited before the IBAD-MgO layer are atomically disordered. Highly textured layers start from the IBAD-MgO layer. Fig. 5 is an HRTEM image of the MgO and LMO layers taken along

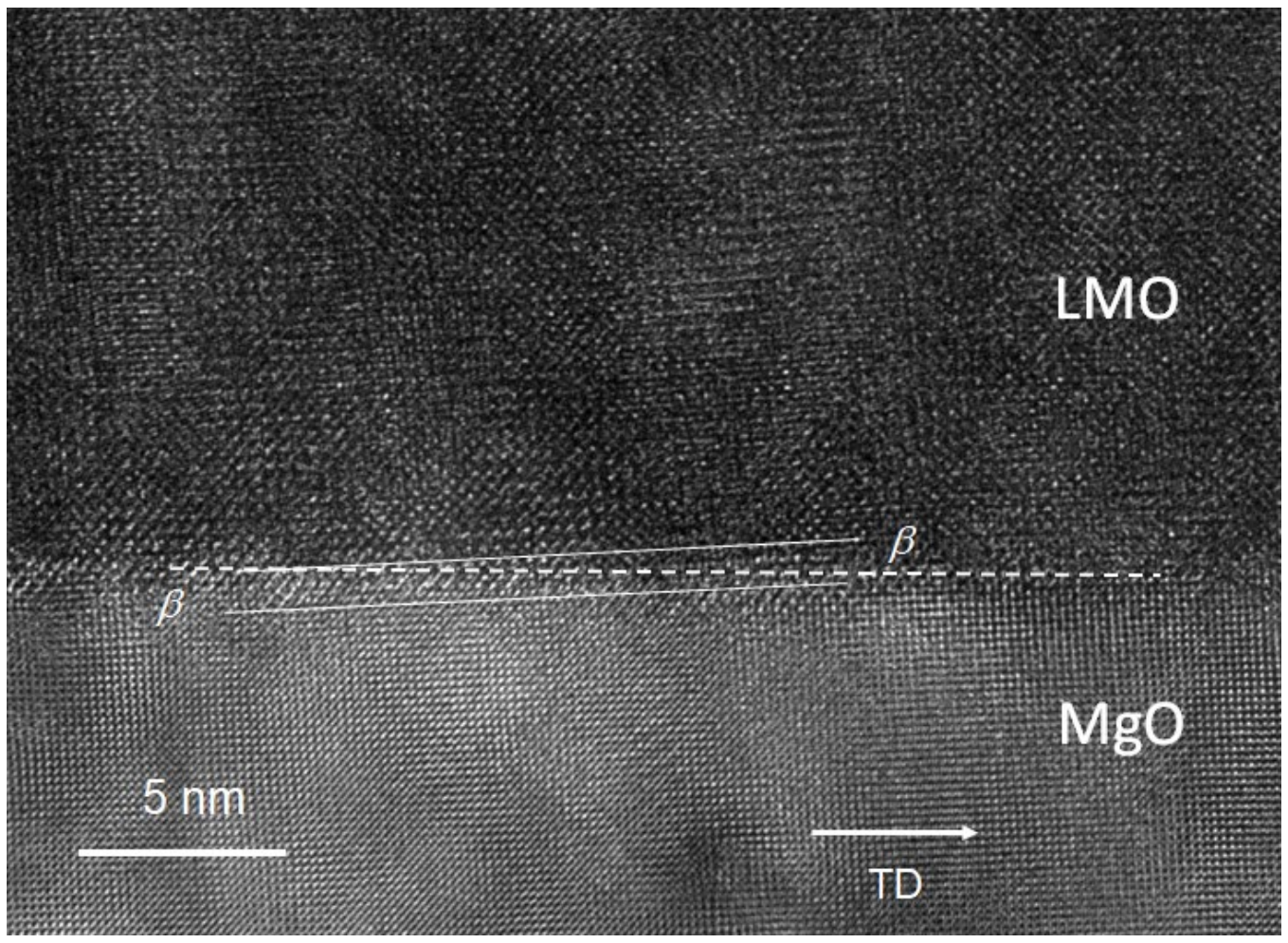

Fig. 5. High resolution TEM image of cross-section of MgO/LMO buffer layer taken along [100] zone axis. The transverse direction of REBCO tape is indicated as 'TD'. The dash line represents the LMO/MgO interface. The solid lines are guides to the eye for atomic planes. The lattice (001) planes is tilted from the interface by $\beta \sim 2.9°$.

[100] zone axis. The transverse direction (TD) of the REBCO tape is indicated in the figure. It is apparent that both MgO (001) and LMO (001) crystal planes are tilted with respect to the MgO/LMO interface which is represented by the dashed line. The tilt angle $\beta$ in Fig. 5 is measured to be 2.9°. This value is consistent, within the measurement error, with XRD result of 2.2° from a sample that was cut from the same production piece length.

### C. *Large volume of tilt angle tests*

As a part of quality assurance test program to serve the 40 T all-superconducting magnet project, *ab*-plane tilt angle was measured for each piece length of incoming SuperPower REBCO tapes. The direction (polarity) of the tilt was determined and documented for each sample. The magnitude of the tilt angle of 68 samples is shown in Fig. 6. The average tilt angle is 2.2° with a standard deviation of 0.5°. Evidently the magnitude of the tilt is nearly constant for these SuperPower REBCO tapes, and the direction (polarity) of the tilt can be correlated with the beam angle of the ion-source with respect to the tape longitudinal growth direction in the IBAD-MgO deposition process.

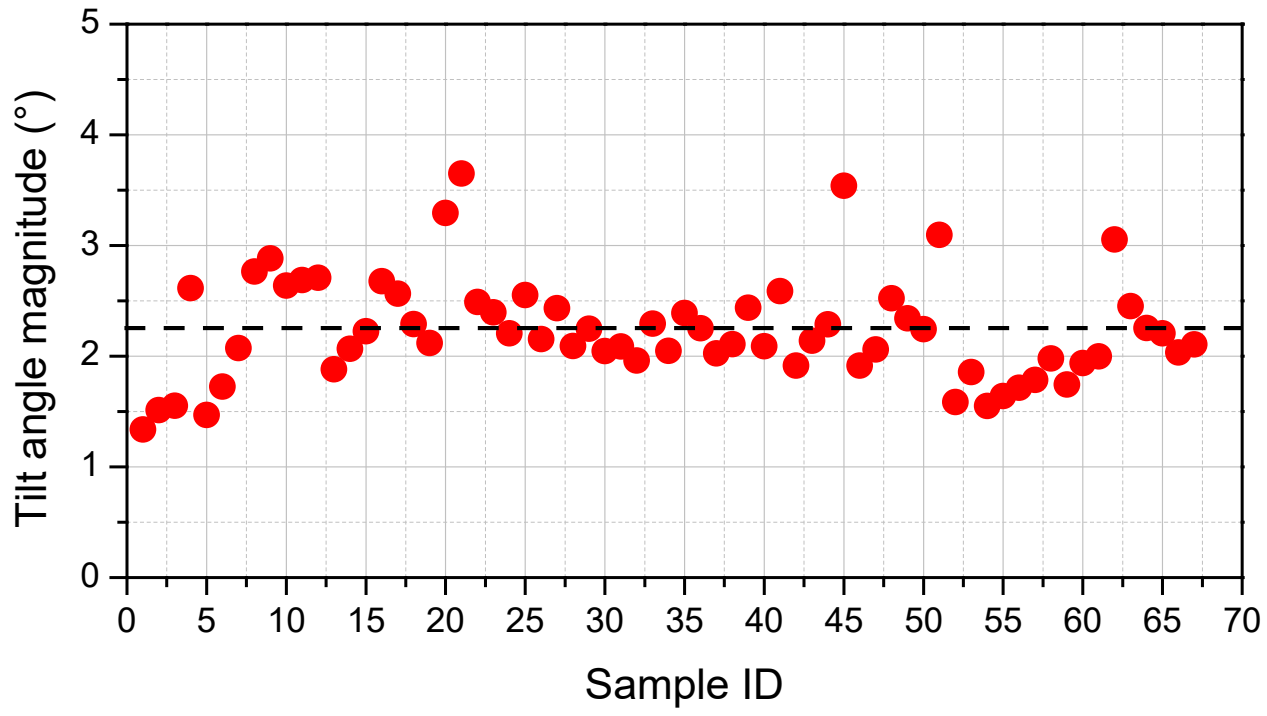

Fig. 6. Magnitude of the tilt angles for 68 samples. The dash line represents the average tilt of 2.2°.

## IV. Discussion

### A. Measurement error in the tilt angle

To evaluate the repeatability of the tilt angle measurement, five samples were cut from a piece of 75 mm long REBCO tape. Each sample was glued to a glass-slide for the tilt angle measurements. Each sample was loaded to the XRD system and measured twice (Test 1 and Test 2). The repeatability test results are shown in Fig. 7. Evidently loading sample to the XRD system (Test 1 versus Test 2) introduced very small variation in tilt angle. However, the scatter between the 5 samples is significant. The standard deviation of the data in Fig. 7 is 0.3° with Max - Min = 0.9°. This scatter can be attributed to the slightly uneven glue between the glass-slide and the sample, which may be reduced in the future by developing a more reliable sample mounting method.

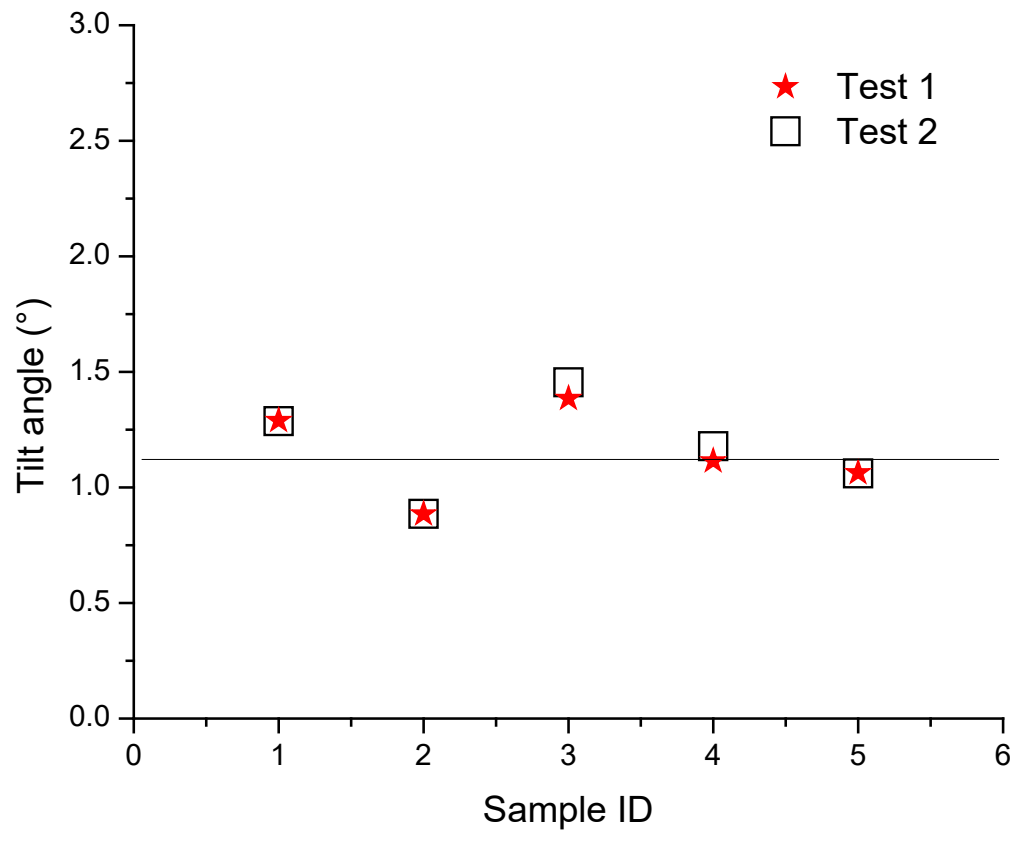

Fig. 7. Repeatability of the *ab*-plane tilt values. Measured on 5 samples cut from a 75 mm conductor. Each sample was tested twice (Test 1 and Test 2).

### B. The origin of the tilt

The origin of the *ab*-plane tilt has been investigated previously [4]-[6]. Reference [4] suggested that in the IBAD process [001] direction of the IBAD-MgO layer is slightly tilt toward the ion beam which was 45° to tape surface in the transverse direction. This 45° ion-beam angle is designed so the ion-beam is along the MgO [110] direction to promote single crystal growth [11]. It is speculated that the small tilt angle is formed naturally to facilitate the vicinal growth (step-flow growth) of the MgO single crystal layer. The reported *ab*-plane tilts from these few groups [4]-[6] suggest that the tilt is a common feature not unique to a specific IBAD growth system, although the value of the tilt angle seems to vary from system to system.

The detailed molecular dynamics of the IBAD process that leads to this tilt is beyond the scope of this work. But if the tilt is always towards the ion-beam, it may be controlled by the configuration of the ion-beam. For example, a rearrangement of the ion-beam to make it 45° inclined in the tape longitudinal direction would result in a tilt in the longitudinal direction. This would result in the same $I_c$ regardless its orientation in transverse direction, which would be desirable for users of REBCO conductors for high field magnet applications.

### C. $I_c(\theta)$ shift due to anisotropy in pinning

It was reported that at liquid nitrogen temperatures and low magnetic fields $I_c(\theta)$ peak shift slightly from *ab*-plane depending on the strength of the applied magnetic field [4]. This is attributed to the anisotropy in pinning by various mechanisms. In high magnetic fields of our interest, however, this shift becomes negligible, which makes $I_c$ peak always at the *ab*-plane. Furthermore, at liquid helium temperatures the pinning seems to be dominated by uncorrelated point pins [12] which are not expected to influence the angle of the peak in $I_c(\theta)$. Ideally, the $I_c(\theta)$ should be measured and compared with XRD tilt angle measurement to verify that $I_c$ is at maximum when B//*ab*-plane. This type of measurements, however, are usually subjected to large electromagnetic force rendering accurate angular measurement (error < 2°) very difficult, especially at very small $\theta$ where $I_c$ and electromagnetic force is near the maximum.

## V. Conclusion

The tilt angle between the REBCO surface plane and its crystallographic *ab*-plane significantly influence $I_c$ in high magnetic fields. We measured the tilt angles for a large number of SuperPower REBCO samples at room temperature by XRD.

It is clearly shown by both XRD and HRTEM that the tilt originates from the IBAD-MgO buffer layer. For total of 68 samples measured, the average tilt angle is 2.2° with a standard deviation of 0.5°. These results have been incorporated in $I_c(\theta)$ database which will be used in the development of the 40 T all-superconducting magnet at the NHMFL.

## Acknowledgment

We thank Prof. Fumitake (Tak) Kametani of Florida State University, College of Engineering for helps in XRD experiments.